\documentclass{article}
\usepackage{geometry}
\usepackage{mathtools}
\usepackage{bm}
\usepackage{amsmath}
\usepackage{braket}
\usepackage{graphicx}
\usepackage{comment}
\usepackage{caption}
\usepackage{subcaption}
\usepackage{hyperref}
\usepackage{amsfonts} 
\usepackage{physics}
\usepackage{dsfont}
\usepackage{multirow}
\usepackage{array}
\newcolumntype{P}[1]{>{\centering\arraybackslash}p{#1}}
\usepackage{multirow}
\usepackage{xcolor}
\usepackage[affil-it]{authblk}
\usepackage[super,comma,sort&compress]{natbib}

\title{Super-resolved CARS by coherent image scanning}
\author[1]{Anna Zhitnitsky}
\author[1]{Elad Benjamin}
\author[2]{Ora  Bitton}
\author[3,*]{Dan Oron}
\date{}

\affil[1]{Weizmann Institute of Science, Department of Physics of Complex Systems, Rehovot, Israel}
\affil[2]{Weizmann Institute of Science, Department of Chemical Research Support, Rehovot, Israel}
\affil[3]{Weizmann Institute of Science, Department of Molecular Chemistry and Materials Science, Rehovot, Israel}
\affil[*]{dan.oron@weizmann.ac.il}

\begin{document}
\maketitle

\section{Abstract}
We present super-resolved coherent anti-Stokes Raman scattering (CARS) microscopy by implementing phase-resolved image scanning  microscopy (ISM), achieving up to two-fold resolution increase as compared with a conventional CARS microscope. Phase-sensitivity is required for the standard pixel-reassignment procedure since the scattered field is coherent, thus the point-spread function (PSF) is well-defined only for the field amplitude. We resolve the complex field by a simple add-on to the CARS setup enabling inline interferometry. Phase-sensitivity offers additional contrast which informs the spatial distribution of both resonant and nonresonant scatterers. As compared with alternative super-resolution schemes in coherent nonlinear microscopy, the proposed method is simple, requires only low-intensity excitation, and is compatible with any conventional forward-detected CARS imaging setup. 

\vspace{10mm}

Far-field super-resolution optical microscopy has advanced from an exploratory concept to a bioimaging reality in the past thirty years. Traditionally, super-resolution modalities have been developed for incoherent signals, predominantly fluorescence. These are based on a variety of physical principles, violating one or more of the underlying assumptions in the derivation of the diffraction limit. Stimulated emission depletion microscopy (STED), which utilizes the extreme nonlinearity of fluorescence depletion (either by stimulated emission or by shelving) is based on quenching fluorescence outside an arbitrarily small volume \cite{Klar1999}. Saturated excitation microscopy (SAX) uses fluorescence saturation directly via observation of the temporal harmonics of the fluorescent signal under modulated excitation \cite{fujita2007}. Techniques like photo-activated localization microscopy (PALM) \cite{Betzig2006} or stochastic optical reconstruction microscopy (STORM) \cite{Rust2006} and super-resolution optical fluctuation imaging (SOFI) \cite{Dertinger2009} rely on the utilization of temporal dynamics in the fluorescence signal. Another approach is based on spatially modulated excitation,  and includes modalities like structured illumination microscopy (SIM) \cite{Gustafsson2000}, Random illumination microscopy (RIM) \cite{mudry2012structured} and image scanning microscopy (ISM) \cite{Enderlein2010}. These methods take advantage of the increased Fourier support due to the spatial modulation of the excitation to improve resolution. While  limited in their ability to increase resolution, these techniques are usually much simpler to implement.

The development of super-resolution for imaging modalities based on coherent scattering is significantly lagging behind that of fluorescence. This is due to several reasons, including: the difficulty to saturate scattering processes, the fact that the point spread function is only defined for the field amplitude (rather than the intensity) in coherent imaging, and the difficulty to temporally modulate the scattering cross section. This is true for all coherent nonlinear scattering based imaging techniques such as second harmonic generation (SHG) \cite{campagnola2001second}, third harmonic generation (THG) \cite{Squier:98}, and coherent nonlinear Raman scattering (including coherent anti-Stokes Raman scattering, CARS \cite{duncan1982scanning}, and stimulated Raman scattering, SRS \cite{freudiger2008label}).
Most of the efforts to perform super-resolution coherent imaging focused on CARS microscopy. CARS is a widely-used vibrational imaging method whereby two laser fields at the pump ($\omega_p$) and Stokes ($\omega_s$) frequencies interact with a medium coherently to generate a new field at the anti-Stokes frequency ($\omega_{as} = 2\omega_p - \omega_s$) via a third-order induced polarization \cite{duncan1982scanning, zumbusch1999three}. CARS therefore affords a non-destructive, label-free method of imaging, which has led it to be one of the most used nonlinear microscopy methods in biology. Nonlinear laser scanning microscopy also offers deeper sample penetration on two fronts: first, the nonlinear intensity dependence provides inherent optical sectioning capability without the use of a pinhole, recovering signal otherwise blocked by the pinhole \cite{denk1990}; second, near infrared (NIR) excitation wavelengths lead to greater penetration depth \cite{kobat2009}. Due to the diffraction limit, this  leads to lower spatial resolution compared to the shorter-wavelength excitation used in fluorescence microscopy. Therefore, enhancing the resolution of CARS microscopy in particular, and other variants of coherent Raman imaging in general, is highly desirable. 

Past attempts to achieve super-resolved coherent Raman microscopy have relied on analogues of fluorescence-based methods. A STED-like configuration was demonstrated by depleting the pump beam using a competing nonlinear process \cite{Silva2016,kim2017selective,Choi2018}. A SAX-like configuration utilized saturation of the vibrational transition \cite{yonemaru2015super, Gong2019}. The use of higher order nonlinearity (that is, $\chi^{(5)}$ or $\chi^{(7)}$ rather than $\chi^{(3)}$ as in standard CARS) \cite{gong2020higher} is, in many senses, equivalent to SAX and serves the same purpose. Nevertheless, achieving saturation or depletion for a third order nonlinear signal (or equivalently the excitation of higher order nonlinearity) typically requires very high excitation powers, at the level of $10^{11} W/cm^{2}$ and significant experimental complexity, making such methods rather impractical for biological microscopy applications. A RIM analogue of CARS imaging was demonstrated utilizing dynamic speckle illumination for both the pump and the Stokes beams in a wide-field CARS microscope \cite{Fantuzzi2023}. This makes the CARS signal quasi-incoherent enabling the use of an analysis similar to that used in fluorescence imaging.

Perhaps the most natural super-resolution modality to be applied to CARS imaging is ISM, which is by definition a laser scanning modality. ISM is a variant of confocal microscopy which offers access to the full signal collected through an open aperture while maintaing the spatial resolution gain from a closed aperture confocal. This is possible through the use of a pixelated detector in place of a bucket detector, where each pixel now behaves as a small pinhole. Proposed  in 1988\cite{sheppard1988}, it was first experimentally demonstrated in 2010 \cite{Enderlein2010}, and it was later applied to SHG microscopy \cite{Gregor2017,rescan} using a similar analysis as in the incoherent case. Yet, it was later shown that for SHG in particular, and for coherent scattering in general, the ISM analysis must take coherence into account and, for that end, requires either an interferometric microscope \cite{Raanan2022} or direct access to the field (as in photoacoustic imaging) \cite{sommer2021}.

In this work we apply ISM to CARS by developing a phase-sensitive CARS microscope that enables reconstruction of the full CARS field and therefore coherent ISM. We show that for CARS there is a spatial variation of phase arising from the relative resonant and nonresonant scattering contributions. This requires knowledge of the full field  in order to reassign pixel intensities appropriately. We show that CARS-ISM achieves a resolution gain of approximately 1.8, which is comparable to all previously reported methods of super-resolved CARS, while operating with significantly lower excitation power than STED- and SAX-like methods and requiring  a relatively simple extension to a standard CARS microscope rather than a specialized setup, unlike other structured illumination approaches like RIM. 

\section{Theory of CARS-ISM}
ISM makes use of a pixelated detector as an array of pinholes in order to retrieve the resolution offered by a closed aperture while making use of signal collected by the entire detector. Due to parallax, each off-axis pixel captures a shifted image, so the straightforward summation of pixel intensities results in lost resolution in the same way as opening a confocal pinhole. Shifting the single-pixel images appropriately, via a process called pixel reassignment, corrects for  parallax, thus regaining the single-pixel resolution while maintaining all of the signal of an open aperture. 

\begin{figure}[h!]
        \centering	\centerline{\includegraphics[scale=0.25]{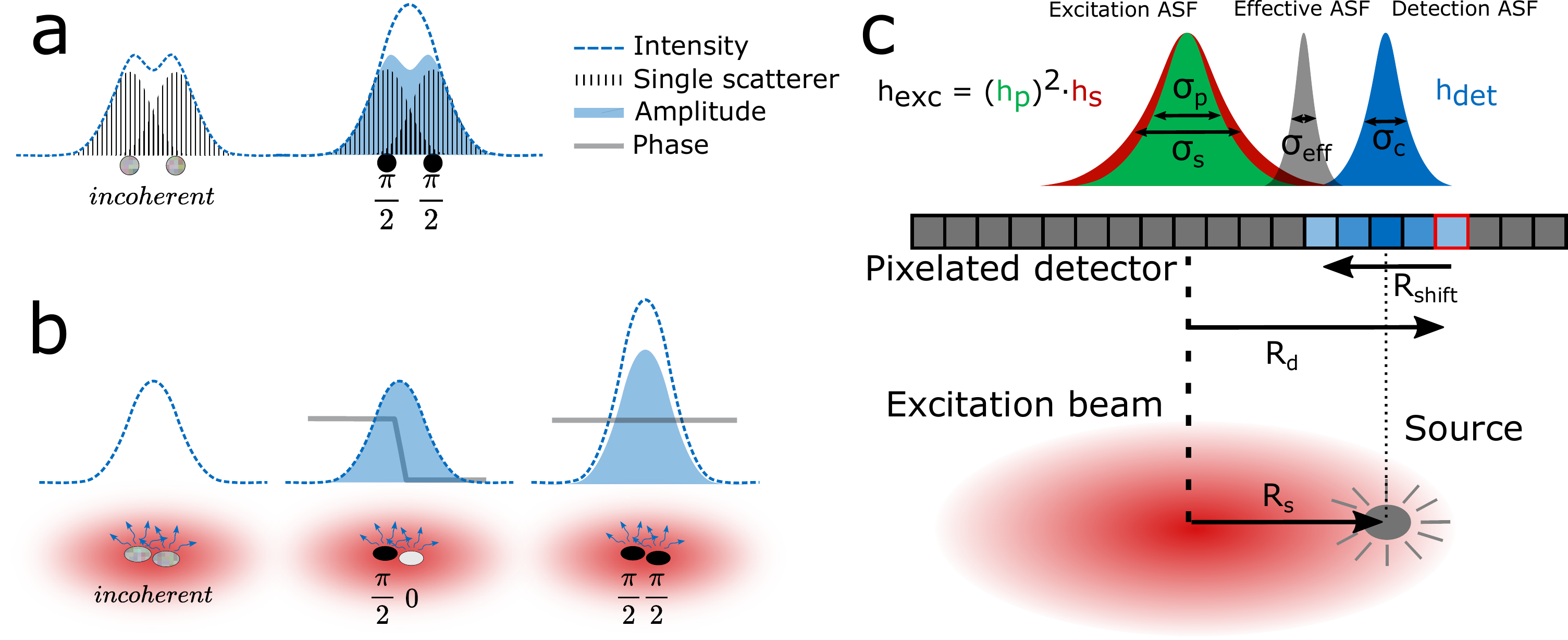}}
        \caption{a) Signal from individual incoherent scatterers  are summed linearly while they must be summed in quadrature for coherent signals meaning pixel reassignment must be done on the amplitude of coherent emitters in order to properly resolve features.   b) The phase impacts the total signal measured on the detector and  having access to phase information distinguishes between resonant and nonresonant signal, which are generated with $\pi/2$ phase difference in CARS.  c) The excitation beam (red) defines the optical axis,  and constitutes the product of the Stokes (dark red) with the squared pump (green) ASF. The detection ASF (blue) represents the response to different locations of the source, assuming flat illumination. The total response (gray) is the multiplication of the excitation and detection ASF, and peaks at a distance described by \autoref{eq:Rshift}}
        \label{fig:cartoon}
    \end{figure}

However, like all PSF-based super-resolution techniques, ISM was developed to enhance incoherent signals, so the pixel reassignment analysis assumes that individual sources do not interfere and therefore that the PSF is well-defined for measured intensity. Yet for CARS, the PSF is only defined for the field amplitude, since even in the simple case of two point sources, the intensity image captured by the camera relates to the fields of two scatterers as follows:
\begin{equation}
        I\left(\textbf{r}\right) \propto |E_1\left(\textbf{r}\right) + E_2\left(\textbf{r}\right)|^2 = I_1\left(\textbf{r}\right) + I_2\left(\textbf{r}\right) + 2\Re\{E_1\left(\textbf{r}\right)E_2^*\left(\textbf{r}\right)\}
    \label{eq:coherent}
\end{equation}
where $E_{1,2}$ are the field amplitudes generated by two point sources. Pixel reassignment must therefore be carried out directly on the field amplitude in order to have access to the full possible resolution and avoid artifacts by considering interference terms, as illustrated in \autoref{fig:cartoon}a and b.  
To compute the necessary pixel reassignment and assess the potential resolution gain from coherent ISM in the CARS context let us  assume spatially invariant and Gaussian excitation and detection amplitude spread functions (ASF). The excitation ASF is proportional to:
    \begin{equation}
        h_{exc}\propto h_p^2h_s\propto \left[e^{-\left(\frac{R_s}{ \sigma_p}\right)^2}\right]^2e^{-\left(\frac{R_s}{ \sigma_s}\right)^2}
        \label{eq:exc}
    \end{equation}
where $h_p,\,  \sigma_p,\, h_s$, and $ \sigma_s$ are the ASF and widths of the pump and Stokes beams, respectively. We define the detection ASF to be the distribution of the detected field of a point source along the camera pixels, given a flat illumination:
    \begin{equation}
        h_{det}\propto e^{-\left(\frac{R_d-R_s}{ \sigma_c}\right)^2}
        \label{eq:det}
    \end{equation}
$R_s$ and $R_d$ represent the location of the emitter and a specific pixel in the array with respect to the optical axis on the sample plane, and $ \sigma_c$ is the width of the CARS signal amplitude spread function. The total response is the multiplication of the excitation and detection amplitude spread functions, as illustrated in \autoref{fig:cartoon}c. We can then calculate the shift in the optical axis needed to compensate for  parallax:
\begin{equation}
    R_{shift} = R_d\frac{ \sigma_p^2 \sigma_s^2}{2 \sigma_c^2 \sigma_s^2 +  \sigma_p^2 \sigma_s^2 +  \sigma_p^2 \sigma_c^2}
    \label{eq:Rshift}
\end{equation}
The estimated resolution from the width of the total response is:
    \begin{equation}
         \sigma_T = \frac{ \sigma_p \sigma_s \sigma_c}{\sqrt{2 \sigma_c^2 \sigma_s^2 +  \sigma_p^2 \sigma_s^2 +  \sigma_p^2 \sigma_c^2}}
        \label{eq:resolution}
    \end{equation}
which is narrower compared to both excitation and detection ASFs, as expected (see Supplementary section 1.1 for a full derivation). In reality, these shifts must be corrected for aberrations. Practically, we obtain the shift vectors numerically by computing the cross-correlation between images from different detector pixels, see Supplementary 1.2. It is worth noting that this approach still assumes PSF shift-invariance. According to \autoref{eq:resolution}, for the experimental conditions used in the experiments described below, we expect to obtain a resolution of $\sigma_T = 220 \text{ nm}$ from coherent ISM. This is compared to the expected resolution given by the excitation ASF: $\sigma_T = \frac{\sigma_p\sigma_s}{\sqrt{2\sigma_s^2+\sigma_p^2}} \approx 326$ nm. Therefore we expect a circa 1.5x resolution gain. Notably, the exact value depends on the exact shape of the ASFs (which are generally not Gaussian). 

In fluorescence-based ISM, Fourier reweighting is often applied to further enhance spatial resolution. This procedure involves amplification of the high spatial frequency components so as to compensate for their reduced value following free-space propagation. Notably, for coherent processes this is significantly less effective since the frequency response is much more uniform over spatial frequencies (see Supplementary Information section 1.4). 

\section{Experimental}
\begin{figure}[htpb]
        \centering	\centerline{\includegraphics[scale=0.075]{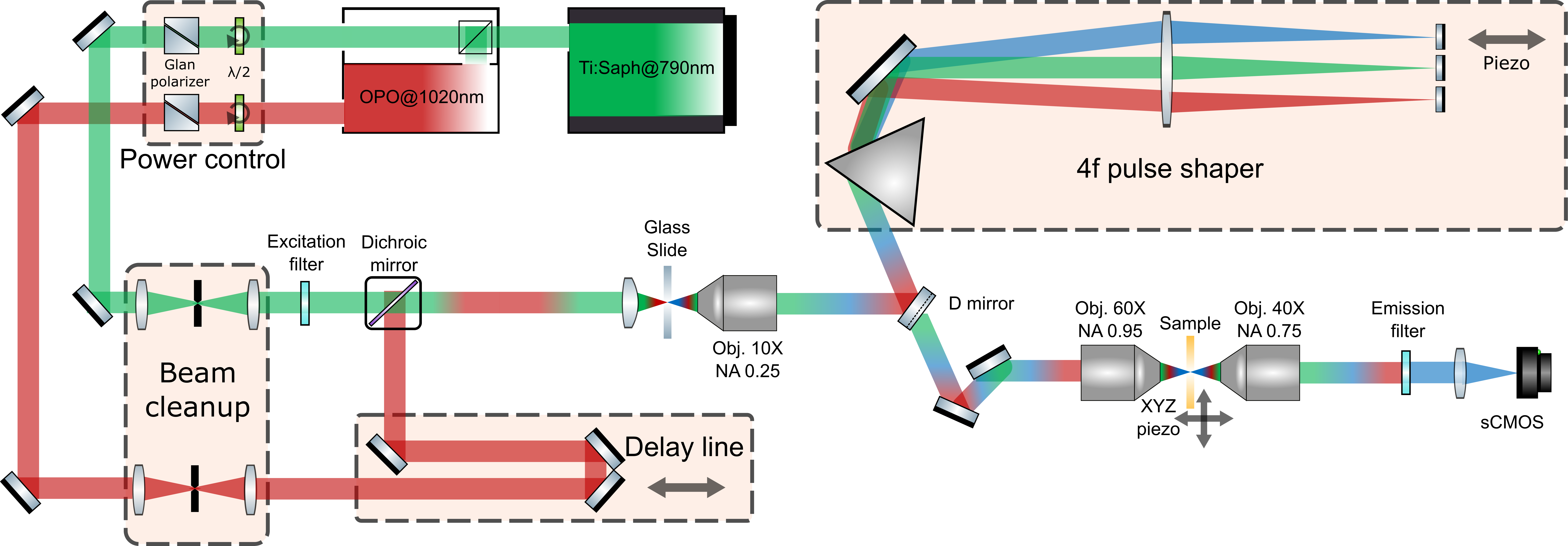}}
        \caption[CARS-CISM setup]{\textbf{CARS-ISM setup} The pump and Stokes laser beams are combined spatially on a dichroic mirror, and the pulses are temporally overlapped by maximizing their sum frequency signal, generated from a BBO crystal. The combined excitation is focused into a glass slide to produce the reference beam. Both excitation and reference are sent into a reflective 4f prism shaper, facilitating scanning of the relative phase between the reference and the CARS signal. The beams are focused and recollected from the sample using high NA objectives. Spectral filters are used to clean the pump beam and to spectrally isolate the CARS signal.}
        \label{fig:simpleCARSsetup}
    \end{figure}

To realize CARS-ISM we must build an interferometrically stable setup for characterizing both the amplitude and the phase of the CARS signal as a function of the sample and detector position. We do this by constructing a nearly inline interferometer where the reference is generated by four-wave mixing in a thin glass slide and propagates collinearly with the pump and Stokes beams, as described schematically in \autoref{fig:simpleCARSsetup}.
Briefly, the {790}{nm} pump beam and the {1020}{nm} Stokes beam are generated by a Ti:sapphire oscillator and a synchronously pumped optical parametric oscillator, respectively. The reference signal is generated via nonresonant four-wave mixing in a glass slide. The phase of the reference beam is externally controlled in a reflective 4f prism-based pulse shaper where a piezoelectric transducer introduces a relative delay between the reference and excitation arms. The pump, Stokes and reference beams are focused onto the sample by a high-NA objective and the forward-scattered signal is collected through a second high-NA objective. The sample is raster scanned with an X-Y-Z piezoelectric stage, and the interference of the reference and CARS signal is spectrally filtered from the excitation and imaged onto a sCMOS camera such that the excitation spot is imaged on an area of circa 10x10 pixels. A detailed description of the setup appears in Supplementary section 2. 

The CARS signal is extracted using phase shifting interferometry, modulating the reference at phases of $\phi = 0,\frac{\pi}{2},\pi,\frac{3\pi}{2}$. The CARS amplitude $\abs{E}$ and phase $\theta$ are then:
\begin{subequations}\label{eq:resolved_field}
\begin{align}
    \lvert E \rvert^2 &= \frac{1}{4}\left[\left(I(0) + I(\pi)\right) \pm \sqrt{\left(I(0) + I(\pi)\right)^2 - \frac{1}{\eta^2}\left[(I(0) - I(\pi))^2 + \left(I\left(\frac{\pi}{2}\right) - I\left(\frac{3\pi}{2}\right)\right)^2\right]}\right] \label{eq:resolved_amp}\\
    \theta &= \arctan\left(-\frac{I\left(\frac{\pi}{2}\right) - I\left(\frac{3\pi}{2}\right)}{I(0) - I(\pi)}\right) \label{eq:resolved_phase}
\end{align}
\end{subequations}
Where $I(\phi)$ is the measured intensity in a specific pixel for a recorded image with the reference beam delayed by phase $\phi$, shown in \autoref{fig:delayIMGS}. For a full derivation see Supplementary 1.3. $\eta$ is a correction compensating for loss of interference contrast introduced by small mismatches in the spectral overlap of the signal and reference pulses, and is typically $\sim0.6-0.7$ in our experiments. We estimate $\eta$ from the intensity interference contrast in the spectral and spatial domain. Once the CARS signal is obtained it can be summed up incoherently over the entire camera to obtain the regular CARS image, pixel-reassigned to obtain the CARS-ISM image and further Fourier-reweighted to obtain the FR-CARS-ISM image. As described above, the pixel reassignment shift map is obtained directly from an experimental calibration to compensate for system and sample aberrations (see Supplementary materials section 1.2).

\section{Results}
\subsection{HEK Cells}
\begin{figure}[htpb]
        \centering	\centerline{\includegraphics[scale=0.8]{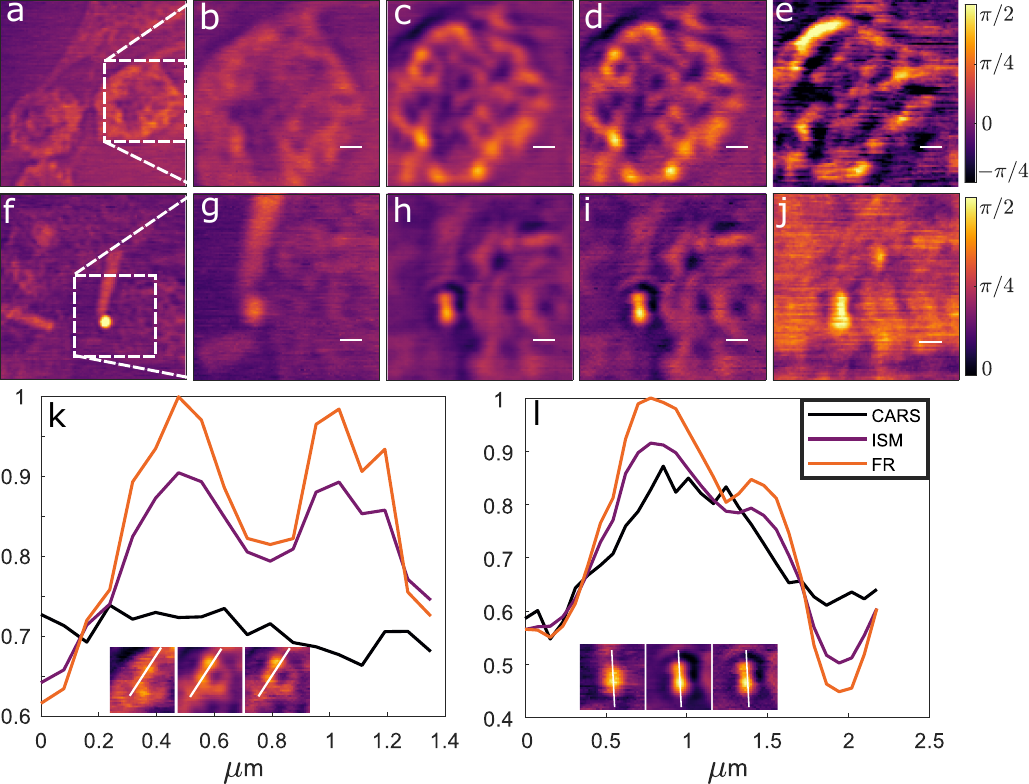}}
        \caption[Comparison]{\textbf{CARS-ISM setup}. CARS intensity images of HEK cells. (a) \& (f) basic CARS image of a large field of view (15x15 $\mu m$) of several cells with squares marking the region imaged in (b)-(e) and (g)-(j) respectively. The intensity $|E_{\text{CARS}}|^2$ is plotted by considering (b) \& (g) straightforward integration of the signal from all detector pixels (basic CARS); (c) \& (h) the ISM image obtained from doing pixel reassignment on the field amplitude captured by each detector pixel; (d) \& (i) the ISM image following Fourier reweighting analysis. All intensity figures are normalized to the same total intensity and plotted on the same colour scale. Additionally the phase images corresponding to the ISM images are shown in (e) \& (j). (k) and (l) show line cuts from the cells in (a)-(e) and (f)-(j) respectively, comparing open-aperture, ISM, and FR-ISM.   Scale bar $= 1 \mu m$.}
        \label{fig:cells}
    \end{figure}

We begin by imaging HEK cells, fixed in Epredia$^{\text{TM}}$  Immu-Mount$^{\text{TM}}$  (Mfr No. Epredia$^{\text{TM}}$ 9990402). Our system is tuned to the characteristic lipid C-H stretch band ($2850 \text{ cm}^{-1}$). To replicate an open-aperture confocal microscope, we treat the camera as a single bucket detector and integrate the signal from all camera pixels. \autoref{fig:cells} compares the CARS intensity ($\abs{E}^2$) images from open-aperture summation of pixels (b \& g), CARS-ISM (c \& h), and FR-CARS-ISM (d \& i). The basic open-aperture CARS images  (b \& g) involve no reference beam in order to provide a fair comparison to a typically-obtained CARS image. Additionally, the phase images shown in (e) and (j) correspond  to the ISM intensity images (c \& h). The spatial variation of the phase conveys the local ratio of resonant and nonresonant scattering, since a resonant CARS signal is generated at a $\pi/2$ phase shift relative to the nonresonant signal. We attribute the experimentally observed negative phases in (e) to phase drifts between the four relative measurements. Our results showcase the additional contrast provided by phase imaging in CARS microscopy. The qualitative similarity between the phase and intensity images indicates the validity of our approach to the reassignment procedure on the field. 

Notably, in contrast with fluorescence microscopy, in CARS imaging we observe a negligible resolution enhancement from Fourier reweighting. This is due to the fact that the magnitude of both the excitation and the detection ASFs is independent of spatial frequency (see Supplementary 1.4 for a details). We nevertheless observe a contrast enhancement from the FR procedure. Line cuts in (k) and (l) show the increased resolution following pixel reassignment, and increased contrast after Fourier reweighting. Analysis of images using the Fourier ring correlation method indicates a resolution increase of 1.5-2x in the CARS-ISM image as compared with the CARS image, and minor (up to 1.1) additional resolution gain following Fourier reweighting (see Supplementary section 1.5 for details). This is in general agreement with the expected resolution increase as derived from equation \autoref{eq:resolution} assuming diffraction-limited excitation beams. 

    \clearpage

\subsection{Resolution}
\begin{figure}[htpb]
 \centering	\centerline{\includegraphics[scale=1]{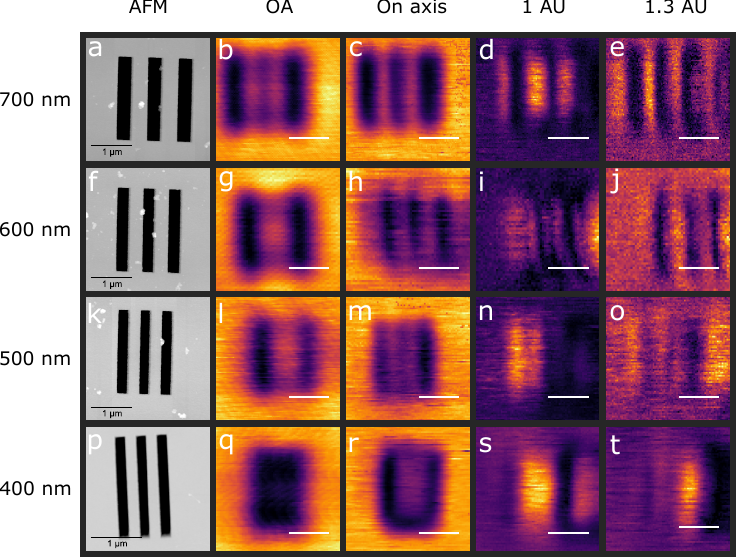}}
    \caption{\textbf{Polymer resolution target.} Ground truth images obtained via AFM are shown in in panels (a), (f), (k), and (p)). (b), (g), (l), and (q) show the standard open aperture (OA) CARS image, where CARS signal (involving no reference beam) from all detector pixels is summed. (c), (h), (m), and (r) show confocal (single on-axis pixel) CARS images, as a proxy for the resolution obtained by pixel reassignment. (d), (i), (n), and (s) show images from pixels  located at 1 Airy unit from the optical axis, and (e), (j), (o), and (t) show images from pixels located at 1.3 Airy units. Scalebar $= 1\mu m$. }
    \label{fig:res_target}
\end{figure}

To provide an alternative quantification of the resolution enhancement obtained using CARS-ISM, we lithographically fabricated resolution targets from a Raman-active polymer (ZEP) as shown in \autoref{fig:res_target}.  Our sample was fabricated using e-beam lithography and characterized by atomic force microscopy (AFM) as described in Supplementary 3. Each element in the target consists of 650nm deep line pairs as shown in \autoref{fig:AFM_profiles}. The periodicity was varied from 400 nm to 700 nm. Since the target consists of a combination of polymer and air gaps, it introduces significant aberrations both in excitation and in detection. We therefore choose to look at the images from several individual detector pixels, mimicking a closed-aperture confocal system, rather than employing pixel reassignment. Images obtained by summation of all detector pixels (in the absence of a reference) are shown in (b), (g), (l), and (q), and represent the basic CARS resolution. The on-axis single-pixel intensity images (squaring the field reconstruction after interference with a reference beam) serve as a proxy for the resolution which can be obtained in CARS-ISM after pixel reassignment, and indicate that the resolving power of our system is improved from circa 700nm in the open-aperture (OA) image to between 400nm and 500 nm, indicating a resolution gain of approximately 1.5-1.6x. Panels (c), (h), (m), and (r) show single-pixel images taken in the first minimum of the detection PSF Airy disk, where we observe significant image distortion due to aberrations induced by the refractive index mismatch of the sample, as discussed in Supplementary 4. Such effects are largely minimized in the images of cells in \autoref{fig:cells}, due to much smaller variation in refractive index. Images from the first Airy ring (e, j, o, and t) show significant distortion but a slightly increased resolution. This is a reasonable outcome of the fact that the system PSF is not fully spatially invariant \cite{rossman2021csparcom}. 
To further supplement our estimate of the resolution gain offered by CARS-ISM, we image an edge in the polymer resolution target and extract via deconvolution a  resolution gain of 1.87 from pixel reassignment (see Supplementary materials 5). The results of FRC analysis, edge deconvolution, and line cuts of biological samples are all in general agreement with the expected resolution gain in the range of 1.5-2.

\subsection{Discussion}
We have shown that the resolution of conventional CARS imaging can be enhanced by a factor of circa 1.8 by a combination of interferometric CARS imaging with ISM and appropriate pixel reassignment (performed on the field amplitude rather than on the intensity). We have also shown that specifically for coherent Raman signals there is an inherent variation of CARS phase on position due to the different relative contributions of resonant and nonresonant scattering, which necessitates the measurement of the complex field amplitude to avoid artifacts. Our technique utilizes a nearly inline interferometer using a modified pulse shaper and as such does not require active stabilization and adds limited complexity to the optical setup while offering a relatively straightforward analysis. The main cost of this is a fourfold increase in the imaging time due to the need to implement phase shifting interferometry, requiring four acquisitions per pixel. Notably, however, in the present realization it can only be practically implemented as forward-scattering CARS.
Different past implementations of super-resolved coherent Raman imaging have utilized somewhat different excitation and collection objectives having different numerical apertures and used different methods for estimating the resolution. Thus, a direct comparison of the obtained spatial resolution may be misleading. It is therefore more useful to compare the resolution gain rather than the absolute values. Practically all previously reported methods exhibit a resolution gain in the range of 1.5-2. For a comprehensive list see \autoref{table:comparison} in the supplementary information file. RIM-CARS showed the highest reported value, of about 2, whereas saturation based methods and higher order CARS showed values of around 1.6. Therefore, the resolution gain from CARS-ISM is comparable to those while not requiring either strong excitation fields (in fact, the use of the reference as a local oscillator enables the detection of relatively weaker signals) or the use of a specialized setup as in widefield RIM-CARS. Regarding the latter, we note that the analysis of RIM-CARS assumes an incoherent signal, while this is not entirely correct at the observed resolution which is similar in length to the size of a single speckle grain.
As presented above, there is still room for improvement of the CARS-ISM setup and sensitivity in various aspects. Due to the use of a reference beam which is focused onto the sample along with the pump and Stokes beams, it is rather sensitive to chromatic aberrations which result in distortion of the reference phase across the image. Further, the phase stability between the reference and the pump and Stokes beams would be greatly improved without the use of three separate reflecting elements in the prism compressor. This could be achieved, for example, by replacing them with a deformable mirror to perform phase shifting interferometry. Such an element could also better compensate dispersion so as to improve the spectral overlap between the reference and the generated signal. At present, the use of a camera limits the acquisition rate at every pixel. Nevertheless, much faster detectors such as monolithic SPAD arrays have already been used in ISM to speed up the acquisition process \cite{castello2019robust}. Exhibiting very low dark count rates (especially upon time gating) these should exhibit similar performance to high-end scientific CMOS cameras.
One should note that the analysis used here utilizes pixel reassignment which makes the approximation of a spatially invariant ASF which is not strictly correct, especially under high numerical aperture illumination and collection (see \autoref{fig:parallax_bead_comparision}). An alternative to pixel reassignment which solves a global optimization problem and avoids this assumption has recently been shown to yield superior results in terms of spatial resolution in fluorescence imaging \cite{rossman2021csparcom}. This method can be adapted to CARS-ISM upon proper calibration of the ASF. 
Finally, we note that many CARS applications require epi-detection of the CARS signal. The phase variation between resonant and nonresonant components of the CARS signal is maintained also for backscattered signals (in contrast with a forward signal which is backscattered following further propagation in the sample). Thus, phase information is still required for proper pixel reassignment. As presented here, the inline interferometry method is not suitable for generating a reference in the backward direction. One possible alternative is to use a strong nonresonant background, generated along with the resonant signal, as a replacement for the reference, but this requires significant modifications to the setup. Another is the use of a wavefront sensor for imaging. This introduces other complexities to CARS-ISM such as the need for phase unwrapping.

\section{Conclusion}
We have presented a simple add-on to a scanning CARS microscope which enables super-resolved CARS by implementing coherent ISM to gain a factor of 1.5-2 in resolution. Coherent ISM relies on appropriately reassigning the field amplitude rather than the intensity, requiring that phase information is retrieved. We measure phase directly by inline interferometry in order to reconstruct the full CARS field. From the phase measurements, we also gain spatially varying information of the samples resonant and nonresonant emitter content. We also suggest that more rigorously accounting for system aberrations, which are enhanced in coherent microscopy, holds potential for further resolution increase. 

\section{Acknowledgements}
We thank Shira Cohen and Professor Philipp Selenko for preparing and providing HEK cell samples. We  thank Irit Goldian for her assistance in obtaining AFM images.

\clearpage
\bibliographystyle{unsrt}
\bibliography{biblio}

\clearpage
\setcounter{figure}{0}
\makeatletter 
\renewcommand{\thefigure}{S\@arabic\c@figure}
\makeatother

\setcounter{equation}{0}
\makeatletter 
\renewcommand{\theequation}{S\@arabic\c@equation}
\makeatother

\setcounter{section}{0}
\section*{Supplementary information}
\section{Image processing}

 \subsection{Derivation of the pixel reassignment shift}
\begin{figure}[h!]
        \centering	\centerline{\includegraphics[scale=0.3]{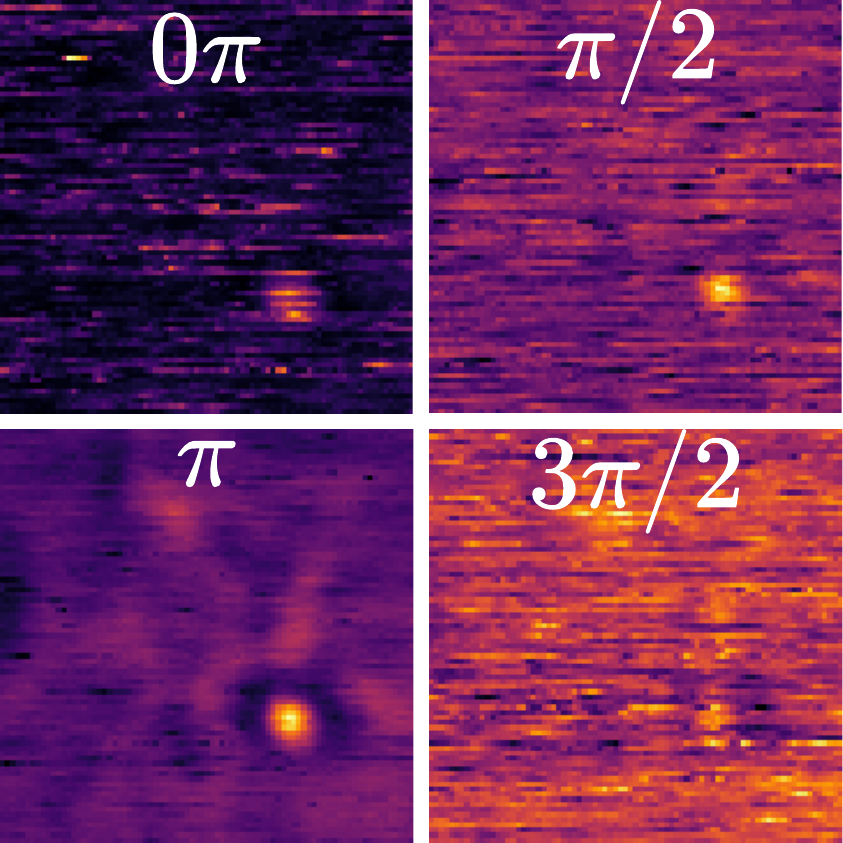}}
         \caption[Interference images]{\textbf{Interference images.} Raw raster-scanned interference images at four different combined phases: $0$, $\pi/2$, $\pi$, $3\pi/2$.}
        \label{fig:delayIMGS}
    \end{figure}

In order to extract the full field we measure the interference pattern of the CARS signal with a reference beam at four different phases. For illustration, the full raster-scanned images obtained from each relative phase are shown in \autoref{fig:delayIMGS} 
 Once the complex-valued amplitude spread functions (ASF) are retrieved for each scan position, we can carry out the pixel-reassignment procedure to shift the ASF, for which the derivation follows. \\
 \\
Assuming the pump and Stokes excitation fields to be Gaussian beams, the excitation amplitude spread function is proportional to the product:
    \begin{equation}
        h_{exc}\propto h_p^2h_s\propto \left[e^{-\left(\frac{R_s}{ \sigma_p}\right)^2}\right]^2e^{-\left(\frac{R_s}{ \sigma_s}\right)^2}= e^{-R_s^2\left(\frac{2}{\sigma_p^2} + \frac{1}{\sigma_s^2}\right)}
        \label{eq:exc}
    \end{equation}
where $h_p,\, \sigma_p,\, h_s,\, and\, \sigma_s$ are the ASF and widths of the pump and Stokes beams respectively. The detection ASF is the distribution of the detected field of a point source along the camera pixels:
\begin{equation}
    h_{det}\propto e^{-\left(\frac{R_d-R_s}{\sigma_c}\right)^2}
    \label{eq:SIdet}
\end{equation}
$R_s$ and $R_d$ represent the location of the emitter and a specific pixel in the array with respect to the optical axis on the sample plane. $\sigma_c$ is the width of the CARS signal ASF. The total response is the multiplication of the excitation and detection amplitude spread functions:
\begin{equation}
    h_{eff} = h_{exc}\cdot h_{det} \propto exp\Biggl\{-R_s^2\left(\frac{2}{\sigma_p^2} + \frac{1}{\sigma_s^2} + \frac{1}{\sigma_c^2}\right) - \frac{R_d^2}{\sigma_c^2} + \frac{2R_sR_d}{\sigma_c^2}\Biggr\}
    \label{eq:SIeff}
\end{equation}
Performing some algebra, \autoref{eq:SIeff} takes the form:
\begin{equation}
    \begin{split}
	h_{\text{eff}} & \propto \exp\Biggl\{-R_s^2\cdot\frac{2\sigma_s^2\sigma_c^2 + \sigma_p^2\sigma_c^2 + \sigma_p^2\sigma_s^2}{\sigma_p^2\sigma_s^2\sigma_c^2} - \frac{R_d^2}{\sigma_c^2} + 2R_sR_d\cdot\frac{\sigma_p^2\sigma_s^2}{\sigma_p^2\sigma_s^2\sigma_c^2}\Biggr\} \\
	& \propto \exp\Biggl\{-\frac{2\sigma_s^2\sigma_c^2 + \sigma_p^2\sigma_c^2 + \sigma_p^2\sigma_s^2}{\sigma_p^2\sigma_s^2\sigma_c^2}\Biggl(R_s^2 - 2R_sR_d\cdot\frac{\sigma_p^2\sigma_s^2}{2\sigma_s^2\sigma_c^2 + \sigma_p^2\sigma_c^2 + \sigma_p^2\sigma_s^2}\Biggr) - \frac{R_d^2}{\sigma_c^2}\Biggr\}
    \end{split}
    \label{eq:SIeffopen}
\end{equation}
We wish to show that $h_{eff}$ is shifted with respect to $R_s$. The expression should therefore contain a Gaussian term $\sim e^{-\left({R_s - R_{shift}}^2\right)}$ where $R_{shift}$ is the shift and should of course depend on $R_d$. Its clear that this would be achieved by ``completing the square" of the term inside the square brackets in \autoref{eq:SIeffopen}:
\begin{multline}
    h_{eff}\propto exp\Biggl\{-\frac{2\sigma_s^2\sigma_c^2 + \sigma_p^2\sigma_c^2 + \sigma_p^2\sigma_s^2}{\sigma_p^2\sigma_s^2\sigma_c^2}\left[R_s - R_d\left(\frac{\sigma_p^2\sigma_s^2}{2\sigma_s^2\sigma_c^2 + \sigma_p^2\sigma_c^2 + \sigma_p^2\sigma_s^2}\right)\right]^2\Biggr\}\\
    \cdot exp\Biggl\{-\frac{R_d^2}{\sigma_c^2}\left(\frac{\sigma_p^2\sigma_s^2}{2\sigma_s^2\sigma_c^2 + \sigma_p^2\sigma_c^2 + \sigma_p^2\sigma_s^2}\right)\Biggr\}
\end{multline}
Hence, the shift is:
\begin{equation}
    R_{shift} = R_d\frac{\sigma_p^2\sigma_s^2}{2\sigma_c^2\sigma_s^2 + \sigma_p^2\sigma_s^2 + \sigma_p^2\sigma_c^2}
    \label{eq:SIshift}
\end{equation}

\subsection{Cross-correlation approach to pixel reassignment}
\begin{figure}[htpb]
    \centering
    \includegraphics[scale=0.5]{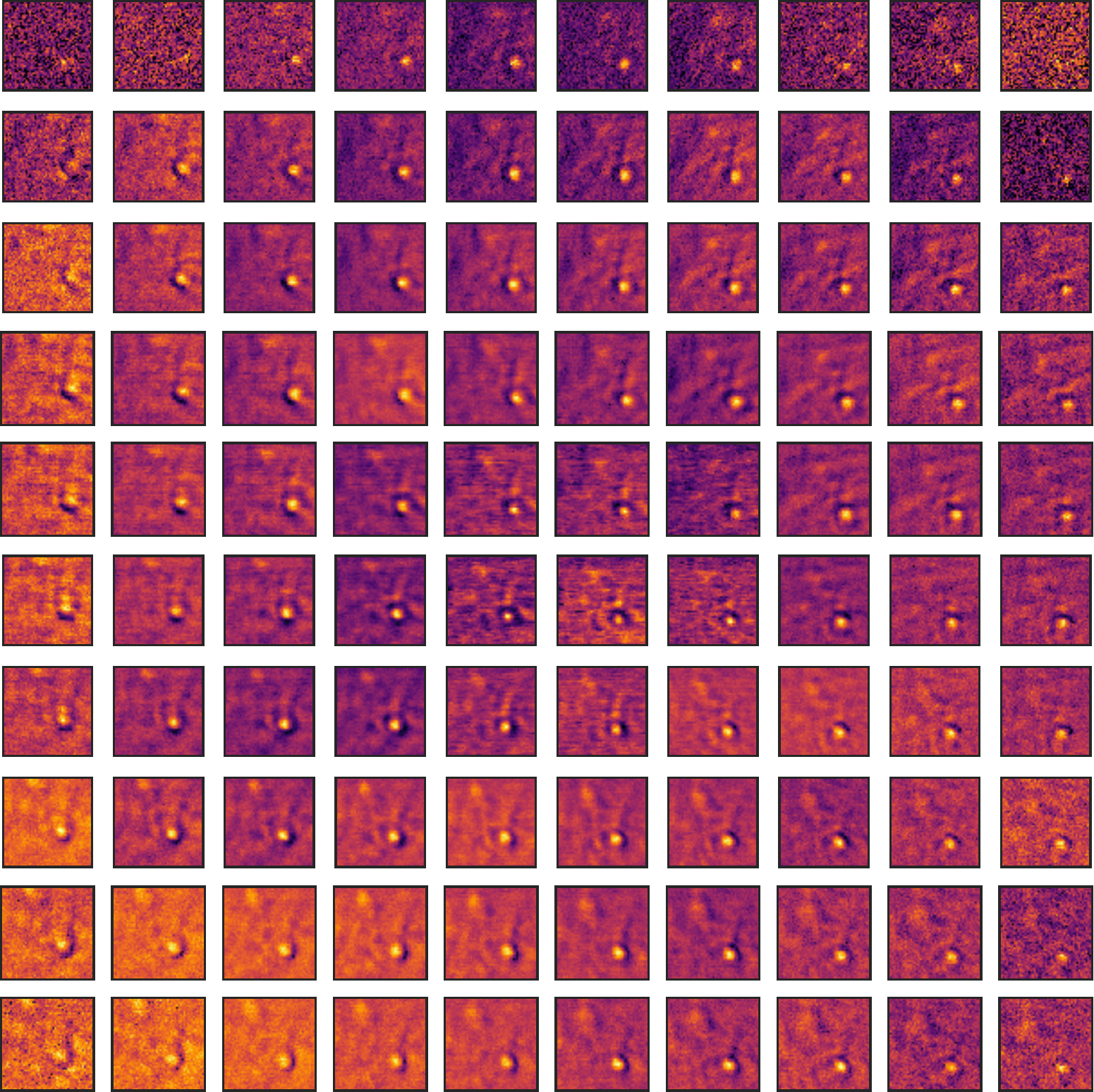}
    \caption{\textbf{Parallax.} Full raster-scanned intensity images of a lipid droplet in a HEK cell, as captured by each pixel on the detector.} 
    \label{fig:pixelIMGS}
\end{figure}

\begin{figure}[htpb]
    \centering
    \includegraphics[scale=0.5]{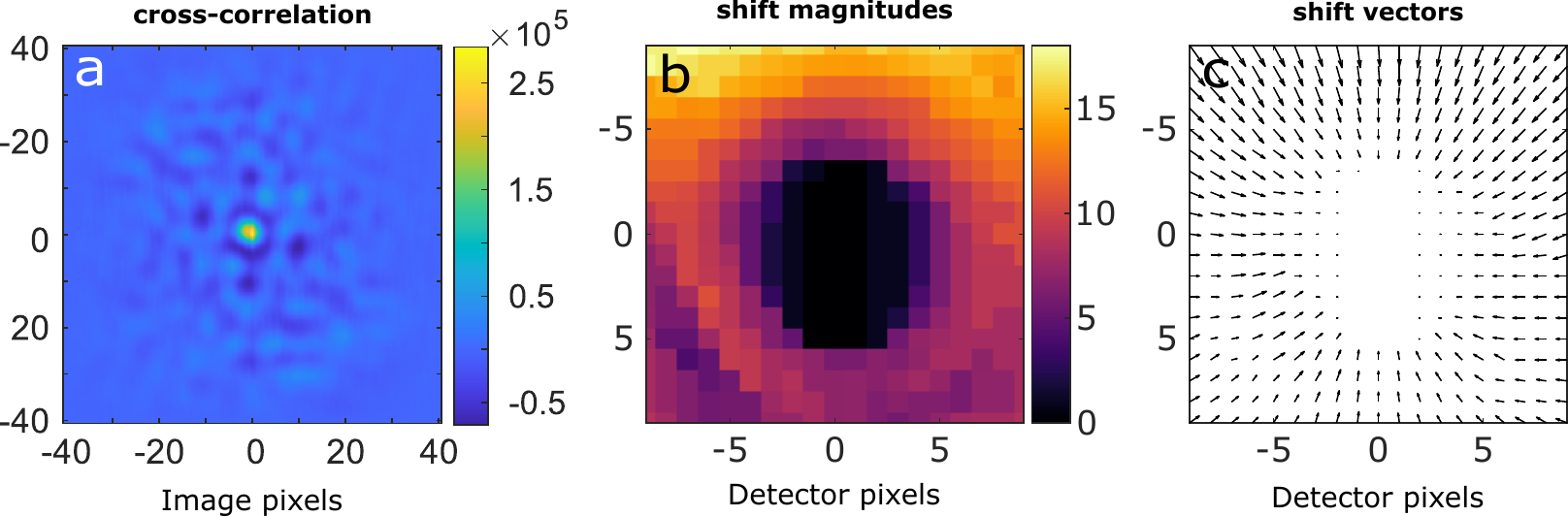}
    \caption{\textbf{Correlation-based image reassignment.} The 2D cross-correlation is computed between the image collected by each detector pixel and the reference (central)  pixel, shown for one such detector pixel in (a). Each image is shifted in accordance with the maximal cross-correlation, resulting in a map of the (b) magnitudes (in pixel number) and (c) vectors of the shifts across the entire detector. }
    \label{fig:correlation_shifts}
\end{figure}

In practice, experimental imperfections such as non-Gaussian beams and spectral abberations cause the required shift to differ from \autoref{eq:SIshift}. An alternative approach to performing the pixel reassignment procedure is by carrying out an image-shifting procedure based on cross-correlation. Each pixel on the camera sees a different image, as shown in \autoref{fig:pixelIMGS}
The image-shifting approach relies on identifying the pixel corresponding to the optical axis (i.e. the closed aperture confocal image), and shifting each other image such that its features overlap with the reference image, thus maximizing signal. The 2D cross-correlation  between each image and the reference is computed (\autoref{fig:correlation_shifts}a, and shifts are assigned according to where the correlation is maximized. The images are interpolated 1x in order to allow for half-pixel shifts.  \autoref{fig:correlation_shifts}b and c show the map of shift magnitude and direction, respectively, across the detector pixels. We see the generally expected behaviour of shifts, albeit with smearing in the y direction due to aberrations of the PSF.

\subsection{Resolving the field and phase (derivation of \autoref{eq:resolved_amp} and \autoref{eq:resolved_phase})}
Start with the general intensity expression in interferometry:
\[ I(\phi) = |E|^2 + |R|^2 + 2\eta |E||R|\cos(\theta + \phi) \]
where \(|R|\) is the amplitude of the reference beam, $\theta$ is the relative phase between the fields and \(\phi\) is the externally controlled phase. Now, let's consider the four intensity measurements:
\[
\begin{aligned}
I(0) &= |E|^2 + |R|^2 + 2\eta |E||R|\cos(\theta) \\
I(\pi/2) &= |E|^2 + |R|^2 - 2\eta |E||R|\sin(\theta) \\
I(\pi) &= |E|^2 + |R|^2 - 2\eta |E||R|\cos(\theta) \\
I(3\pi/2) &= |E|^2 + |R|^2 + 2\eta |E||R|\sin(\theta)
\end{aligned}
\]
Now, let's manipulate these equations to find an expression for \(|E|\). The intensity measurements at different phase shifts are related to the field amplitude \(|E|\) and reference amplitude \(|R|\) as follows:
\begin{align}
I(0) + I(\pi) &= 2|E|^2 + 2|R|^2 \\
I(0) - I(\pi) &= 4 \eta |E||R| \cos(\theta) \\
I\left(\frac{\pi}{2}\right) - I\left(\frac{3\pi}{2}\right) &= -4\eta|E||R| \sin(\theta)
\end{align}
Here we see that the field phase \(\theta\) can be determined from the phase difference between the measurements \(I\left(\frac{\pi}{2}\right)\) and \(I\left(\frac{3\pi}{2}\right)\) relative to \(I(0)\) and \(I(\pi)\):
\[
\theta = \arctan\left(-\frac{I\left(\frac{\pi}{2}\right) - I\left(\frac{3\pi}{2}\right)}{I(0) - I(\pi)} \right)= \arctan\left( \frac{-4\eta|E||R|\sin(\theta)}{ 4 \eta |E||R| \cos(\theta)}\right)
= \arctan(\tan (\theta))\]
Solving the second equation for \(|E|\), we get:
\[
|E| = \frac{1}{4\eta|R|\cos(\theta)} \left( I(0) - I(\pi) \right)
\]
Substituting this into the first equation:
\[
I(0) + I(\pi) = 2\left( \frac{1}{4\eta|R|\cos(\theta)} \left( I(0) - I(\pi) \right) \right)^2 + 2|R|^2
\]
Solving for \(|R|\):
\[
|R|^2 = \frac{1}{4}\left[ I(0) + I(\pi) \pm \sqrt{(I(0) + I(\pi))^2 - \frac{1}{\eta^2}\left((I(0) - I(\pi))^2 + \left(I\left(\frac{\pi}{2}\right) - I\left(\frac{3\pi}{2}\right)\right)^2\right)} \right]
\]
We choose the appropriate solution based on physical considerations. Now, substitute the expression for \(|R|\) into the equation for \(|E|\):
\[
|E| = \frac{1}{4}\left( I(0) - I(\pi) \right) \left[ I(0) + I(\pi) \pm \sqrt{(I(0) + I(\pi))^2 - \frac{1}{\eta^2}\left((I(0) - I(\pi))^2 + \left(I\left(\frac{\pi}{2}\right) - I\left(\frac{3\pi}{2}\right)\right)^2\right)} \right]
\]
This expression gives \(|E|\) explicitly in terms of the measured intensities and the interference contrast \(\eta\). This derivation assumes that the reference beam has a constant amplitude \(|R|\).

\subsection{Fourier reweighting}
Fourier reweighting is a fundamental technique in optical image processing aimed at selectively enhancing or suppressing specific spatial frequencies within an image.  Fourier reweighting is accomplished using the optical transfer function (OTF). The output field $E_{FR}$ can be expressed as $E_{FR}(x, y) = \mathcal{F}^{-1} \left[ \mathcal{F}[E_{ISM}(x, y)] \cdot H(u, v) \right]$, where $E_{ISM}(x, y)$ represents the  input field, $\mathcal{F}$ denotes the Fourier transform operation, $\mathcal{F}^{-1}$ represents the inverse Fourier transform operation, $H(u, v)$ signifies the optical transfer function (OTF) in the frequency domain at spatial frequency coordinates $(u, v)$, and $(x, y)$ are the spatial coordinates. This operation enhances higher frequency components, leading to sharpening edges and highlighting fine details.  We assumed a Gaussian optical transfer function, the width of which was estimated by measuring a 300 nm polystyrene bead. However, because 300 nm is not sufficiently small relative to the resolution limit of the system, and smaller beads do not produce enough signal for a reliable measurement, we fine-tuned the width of the PSF numerically to produce the best contrast in the Fourier reweighted images. 

In coherent scanning optical configurations, the emitted field at some point $(\textbf{r}_d)$ in the vicinity of the detector, from a specific scanning location $(\textbf{r}_s)$ is:
\begin{equation}
    U(\textbf{r}_s) \propto \iint d\textbf{r}_dd\textbf{r}_oh_d(\textbf{r}_d-\textbf{r}_o)\cdot h_{ex}(\textbf{r}_o)\cdot C(\textbf{r}_o + \textbf{r}_s)
\end{equation}
where $\textbf{r}_o$ denotes a point in the vicinity of the sample plane, relative to the optical axis, and $C(\textbf{r}_o)$ is the concentration of emitting material at that point. $h_{ex}(\textbf{r}_o)$ is the excitation amplitude impulse response or \textit{Amplitude Spread Function (ASF)} and $h_d(\textbf{r}_d-\textbf{r}_o)$ is the detection ASF. We take $h_d$ to be a generalization of a lens (where the magnification $M$ is incorporated into the \textit{reduced coordinates} $(\textbf{r}_d)$). Under a few reasonable assumptions, in the Fresnel approximation the detection ASF is just the Fourier transform of the exit pupil of the optical system.
Indeed, we can define an effective ASF:
\begin{equation}\label{eq:effectivePSF}
    h_{eff}(\textbf{r}_o) = \int d\textbf{r}_dh_d(\textbf{r}_d-\textbf{r}_o)\cdot h_{ex}(\textbf{r}_o)
\end{equation}
so that the image can be defined as a convolution between $C$ and said effective PSF:
\begin{equation}\label{eq:stdConv}
    U(\textbf{r}_s) \propto \int d\textbf{r}_oh_{eff}(\textbf{r}_o)\cdot C(\textbf{r}_o + \textbf{r}_s)
\end{equation}
The frequency support is given by the Fourier transform of the ASF - the \textit{Amplitude transfer function (ATF)}. A non-uniform ATF, i.e. some frequency components (lower than the cutoff frequency $\abs{\textbf{k}_c}$) are transferred worse than others, will result in loss of imaging contrast and the observed resolution will be less than $1/\abs{\textbf{k}_c}$. In such cases some contrast can be restored by appropriately offsetting the functional shape of the ATF in the image frequency spectra, a process termed \textit{Fourier re-weighting (FR)}.
Applying the convolution theorem, \autoref{eq:effectivePSF} becomes:
\begin{equation}
    H_{eff}(\textbf{k}) = H_{d}(\textbf{k})\cdot H_{ex}(\textbf{k})
\end{equation}
$H_d$ should have the same functional shape as the exit pupil of the optical system, i.e. uniform up-to a cutoff at $\abs{\textbf{k}_c} = 2NA_d/\lambda$. Hence, assuming a diffraction-limited Gaussian $h_{ex}$, the effective ATF will be of the form:
\begin{equation}
    H_{eff}(\textbf{k})\sim exp\left(-\frac{\abs{\textbf{k}}^2}{2\abs{\textbf{k}_c^{(ex)}}}\right)\cdot circ\left(\frac{\abs{\textbf{k}}}{\abs{\textbf{k}_c^{(d)}}}\right)
\end{equation}
Which is only somewhat non-uniform due to the Gaussian "hump" over the circ function. Hence preforming FR will result in only a modest resolution improvement. This in contrast to the incoherent equivalent, where the detection \textit{optical transfer function} which is an autocorrelation of the coherent ATF and thus triangular with double the cutoff frequency, upon which preforming FR is expected to improve the observed resolution by a factor of $\sim \sqrt{2}$.

\subsection{Fourier ring correlation}
In order to quantify the resolution increase from pixel reassignment, we use  Fourier ring correlation (FRC). FRC is a widely utilized technique in image analysis to assess the resolution and quality of reconstructed images. It involves computing the correlation between two copies of Fourier transformed images obtained from the same sample. The FRC curve provides  information about the spatial frequency at which the imaging system can reliably resolve features, thus serving as a quantitative measure of image quality and resolution. 

The absolute value for resolution provided by Fourier Ring Correlation (FRC) analysis is not entirely reliable due to several factors inherent in the technique and its underlying assumptions. Firstly, FRC relies on the assumption that the two copies of the image being compared differ only in noise while maintaining identical underlying features. However, in practice, variations in  imaging conditions and other experimental factors can introduce additional differences beyond noise, leading to inaccuracies in the resolution estimation. Secondly, the FRC method is sensitive to the choice of parameters such as ring sizes and noise levels. Suboptimal selection of these parameters can result in biased or misleading FRC curves, impacting the accuracy of the resolution measurement. Furthermore, FRC  does not account for other aspects of image quality such as contrast, artifacts, and aberrations that can influence the overall image fidelity. Therefore, while FRC analysis is a valuable tool for assessing relative changes in resolution and comparing different imaging systems or conditions, its absolute resolution values should be interpreted cautiously and validated using complementary techniques or ground truth references to ensure robust and reliable results. We therefore use the results from FRC only to evaluate the relative improvement in resolution offered by pixel reassignment and Fourier reweighting. Importantly, FRC analysis is performed on the final intensity images, not amplitudes.

We use the half bit thresholding method in the resolution analysis. The choice of the half bit method was motivated by the moderate signal-to-noise ratio (SNR) observed in our experimental data, which posed challenges for other thresholding methods such as the 1/7 thresholding method and the three sigma thresholding method. The 1/7 thresholding method is considered less suitable due to its sensitivity to noise. This method has a tendency to overestimate resolution, particularly in datasets with varying noise levels, as it employs a fixed threshold value that may not effectively distinguish signal from noise. The three sigma thresholding method assumes a normal noise distribution and stationary noise, neither of which we can assume in our system due to various experimental factors like periodically drifting laser in intensity. Despite its statistical robustness, this method also requires a relatively high SNR to provide accurate resolution estimates and may be too conservative for datasets with complex noise characteristics or non-normal noise distributions. In contrast, the half bit thresholding method strikes a balance between sensitivity to signal features and robustness against noise, making it somewhat more suitable for estimating resolution increase in our images with varying noise levels. Additionally, we observe that half-bit thresholding provides the most consistent results across various levels of interpolation, meaning that it is the least susceptible to noise. We present FRC results involving 1x interpolation in order to have sufficient pixels to plot smoothed FRC curves. 

\begin{figure}[htpb]
        \centering	\centerline{\includegraphics[scale=0.5]{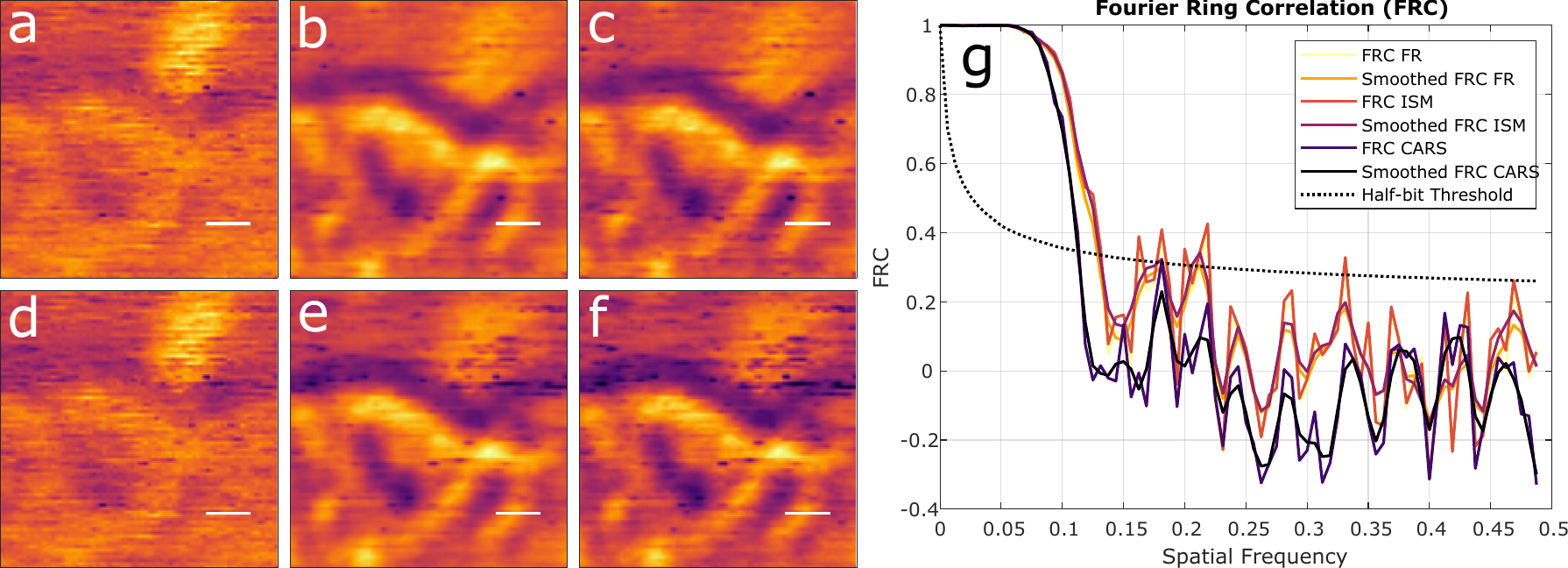}}
        \caption{\textbf{Fourier Ring Correlation.} Two copies of CARS images used for evaluating resolution with Fourier ring correlation: (a) \& (d) basic CARS, (b) \& (e) CARS-ISM, and (c) \& (f) FR-CARS-ISM. The Fourier ring correlations are plotted in (g), showing threshold crossings at 334 nm (CARS), 212 nm (ISM-CARS), and 178 nm (FR-ISM-CARS). The images are 1x interpolated in order to smoothen the FRC curves and more reliably determine the threshold crossing. Scalebar $= 1 \mu m$. }
        \label{fig:FRC}
    \end{figure}

\autoref{fig:FRC}g shows the FRC curves between the two copies of images (a)-(c) and (d)-(f), with a half-bit threshold. Specifically, they show that pixel reassignment improves the resolution by a factor of at least 1.5 (up to 2x), while Fourier reweighting offers small resolution enhancements (between 0 and 0.1x in addition to to CARS-ISM). \autoref{table:FRC-summary} summarizes the results of FRC analysis on the image in \autoref{fig:FRC} to show the large variance in  resolution estimate as a result of different thresholding methods. Nevertheless, the results show that the resolution gain from CARS-ISM falls in the range of 1.5-2.  We note that the half-bit threshold is the most robust to interpolation.

    \begin{table}
\begin{tabular}{ |P{0.3cm}|P{1.8cm}||P{1.8cm}|P{1.8cm}|P{1.8cm}||P{2.1cm}|P{2.1cm}| }
 \hline
&& CARS & ISM & FR & \textbf{CARS/ISM} & ISM/FR   \\
 \hline
\multirow{3}{0.3cm}{\rotatebox[origin=c]{90}{0x}} & 1/7 & 308 & 194 & 241 & \textbf{1.6} & 0.8   \\
 \cline{2-7}
&half-bit & 326 & 224 & 213 & \textbf{1.5} & 1.1   \\
 \cline{2-7}
&three sigma & 364 & 278 & 314 & \textbf{1.3} & 0.9   \\
 \hline
\multirow{3}{0.3cm}{\rotatebox[origin=c]{90}{1x}} & 1/7 & 318 & 168 & 168 &\textbf{1.9} & 1   \\
 \cline{2-7}
&half-bit & 335 & 230 & 223 & \textbf{1.5} & 1\\
 \cline{2-7}
&three sigma & 358 & 301 & 311 & \textbf{1.2} & 1   \\
 \hline
 \end{tabular}
 \caption{Results of FRC analysis for different thresholds, both with and without interpolation. Resolution given in nm.}
\label{table:FRC-summary}
\end{table}

\section{Detailed description of the setup}
   \begin{figure}[htpb]
        \centering	\centerline{\includegraphics[scale=0.08]{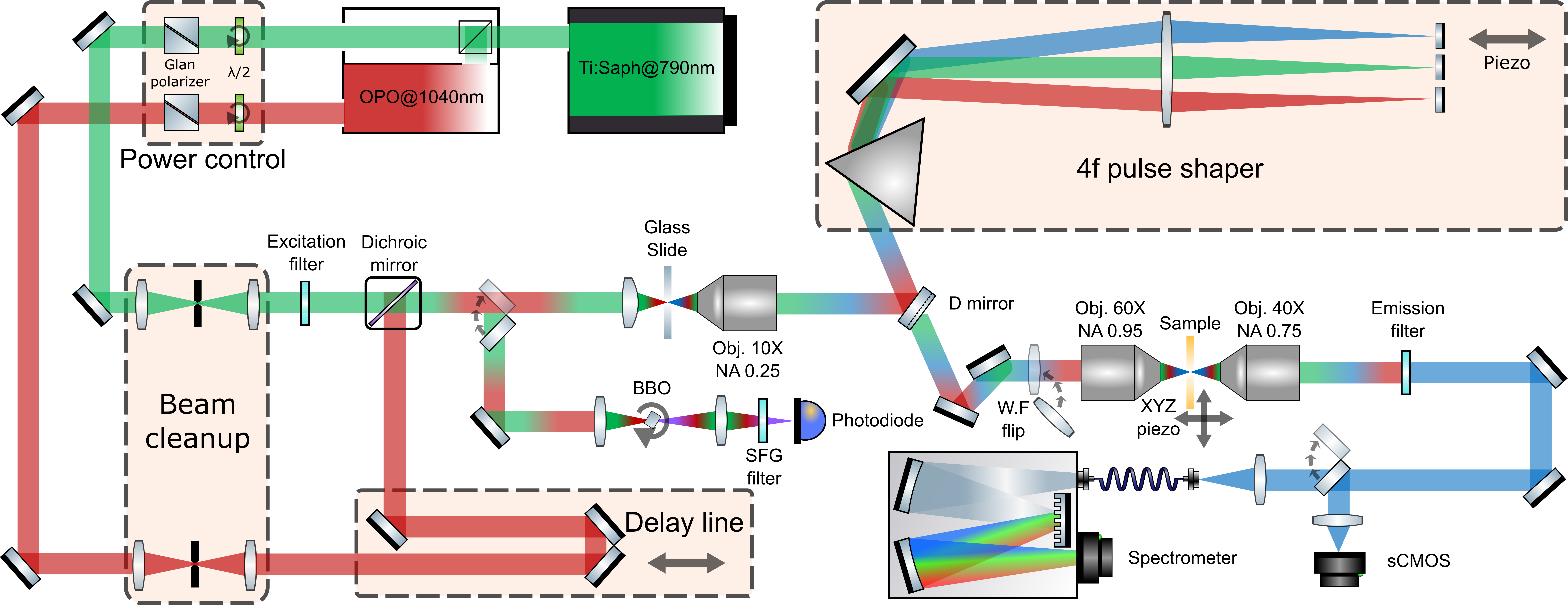}}
        \caption[CARS-CISM Experimental apparatus]{\textbf{CARS-CISM Experimental apparatus.} The pump and Stokes laser beams are combined spatially on a dichroic mirror, and the pulses are temporally overlapped by maximizing their sum frequency signal, generated from a BBO crystal. The combined excitation is focused into a a glass slide to produce the reference beam. Both excitation and reference are sent into a 4f prism shaper, where the relative phase scanning between the reference and CARS signal is realized. The beams are focused and recollected from the sample using high NA objectives. Spectral filters are used to clean the pump beam and to spectrally isolate the CARS signal.}
        \label{fig:CARSsetup}
    \end{figure}
Our phase resolved CARS microscope is illustrated in \autoref{fig:CARSsetup}. A 790 nm pump beam is generated from an ultrafast Ti:sapphire mode-locked oscillator (Chameleon Ultra II, Coherent). A part of the pump is directed into an optical parametric oscillator (Chameleon Compact OPO VIS Version 1.2, Coherent) to generate a 1020 nm Stokes beam. The pulse width for both beams is approximately 150 fs and the repetition rate is 80 MHz. Both beams are passed through a half wave plate - polarizer power control module, and a 4f pinhole arrangement to produce clean single-mode Gaussian beams. The Stokes path is varied through a retro-reflector on a motorized linear stage in order to temporally overlap the pump and Stokes pulses. The beams are then combined at a 890 nm short-pass dichroic mirror. The temporal overlap of the pump and Stokes beams is done by focusing the combined beams into a type I BBO crystal and maximising the SFG signal. The combined excitation is focused into a glass slide to generate a reference signal, through nonresonant 4-wave mixing, and re-collimated along with the excitation using a small achromatic objective lens (Plan N 10X 0.25NA, Olympus). Phase sensitivity is achieved by directing the combined pump, Stokes and reference beams into a folded prism based, 4f pulse shaper, where we correct any group delay between the pump and Stokes and scan the relative phase between the excitation and reference using a piezoelectric transducer on the reference arm. Finally the beams are focused onto the sample using an achromatic objective lens (Plan Apo $\lambda$ 60X 0.95NA, Nikon). Forward propagating signal is collected with another objective lens (Plan-Neofluar 40X 0.75NA, Zeiss), spectrally filtered from the excitation, and focused onto a low dark counts sCMOS camera (pco.edge 5.5, PCO). The sample is mounted on an X-Y-Z piezoelectric stage (P-853 micrometers with E-663 amplifier, PI). Spectra can be measured at various points along the optical path for analysis and alignment (Shamrock 303i with iDus 420, Andor). \\

\section{Sample preparation and characterisation}
\subsection{Electron-beam lithography}
The lines were patterned by electron beam lithography technique, using the e-Line Plus (RAITH) system. The glass substrate was spin coated  with ZEP 520A resist at 1500 rpm, 2000 rpm, 3000 rpm, yielding a thickness of 650 nm, 500 nm, and 400 nm respectively. The sample was then baked at 180 $^o$C for 3 minutes. In order to avoid charging, a 20 nm thick  gold layer was deposited on top of the resist  by thermal evaporation technique.  30 KeV acceleration voltage and 10 $\mu$m aperture size were used, resulting in a beam current of 36 pAmp. The lines were exposed at a dose of 50 $\mu \text{C}/\text{cm}^2$ with step size of 6 nm to yield the desired linewidth.  After the exposure, the sample was developed in Amyl-Acetate for 1 minute followed by immersing in isopropanol for 1 minute. 

\subsection{AFM}
\begin{figure}[h!]
        \centering	\centerline{\includegraphics[scale=0.75]{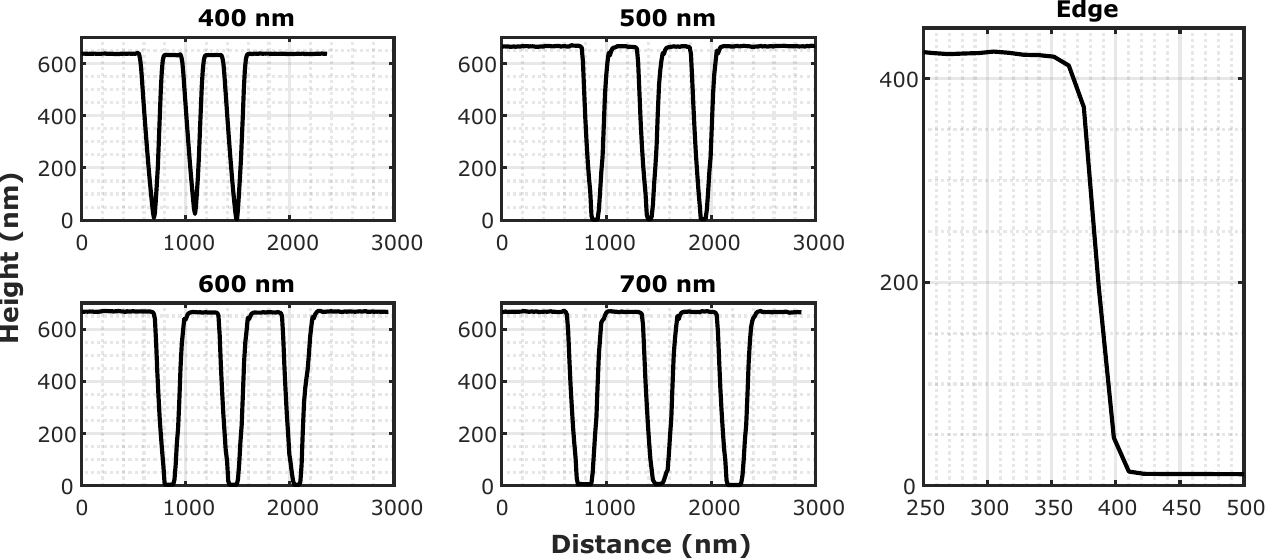}}
        \caption{\textbf{AFM profiles.} }
        \label{fig:AFM_profiles}
    \end{figure}

The polymer resolution targets where characterized using atomic force microscopy (AFM). An OPUS 240AC-NA cantilever tip was used in soft tapping mode on an  Omegascope R SPM (AIST-NT, Horiba). Line-cut profiles of the resolution targets are shown in figure \autoref{fig:AFM_profiles} panels 1-4. An edge was also measured in order to establish the sharpness of the resolution target edges. We observe that the slope of the measured edge is approximately 407 over 59 nm, corresponding to an angle of $\tan^{-1}({\frac{407}{59}}) \approx 81.7^o$. This precisely corresponds to the slope expected from the side half cone angle of the 240AC-NA tip, which contributes $<9^o$, indicating that the resolution of the AFM measurements is tip-limited. We therefore assume that the edges of the resolution target are sufficiently sharp for the purpose of evaluating optical resolution.  

\section{Distribution of phase information}
We observed varying intensity and phase quality across the detector array, shown in  \autoref{fig:parallax_comparision} a and b. Notably, phase images exhibit better quality on the periphery of the airy disk,  beyond 1 Airy unit, compared to the centre of the Airy disk, while intensity images appeared sharper yet noisier and more abberated in the peripheral regions. We attribute the increased resolution  to the distribution of higher frequency components. This behavior is reminiscent of oblique illumination techniques, where off-axis illumination enhances certain features at the cost of introducing artifacts or noise. The distribution of phase and intensity information across the detector array provides valuable insights for implementing  pixel reassignment. By leveraging regions with better phase quality, addressing intensity-noise trade-offs, and optimizing ISM parameters, we can enhance the quality and reliability of phase and intensity reconstructions. Importantly, this requires us to  consider the trade-off between over-sampling intensity and under-sampling phase.  \autoref{fig:parallax_comparision}c shows a similar figure of intensity distribution but involving no reference beam (i.e. basic CARS imaging). We observe that nearly all of the deformations observed on the detector are related to chromatic aberrations of the reference beam. 

\begin{figure}[h!]
     \centering
     \begin{subfigure}[b]{0.9\textwidth}
         \centering
         \includegraphics[width=\textwidth]{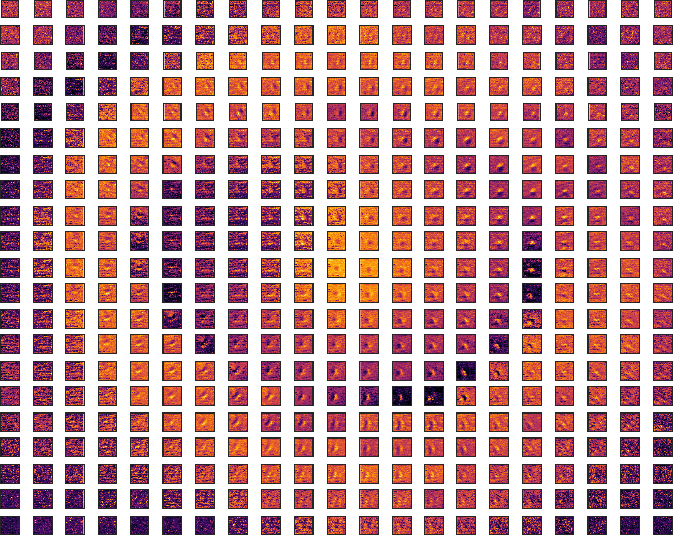}
         \caption{}
         \label{fig:phase}
     \end{subfigure}
     \hfill
     \begin{subfigure}[b]{0.9\textwidth}
         \centering
         \includegraphics[width=\textwidth]{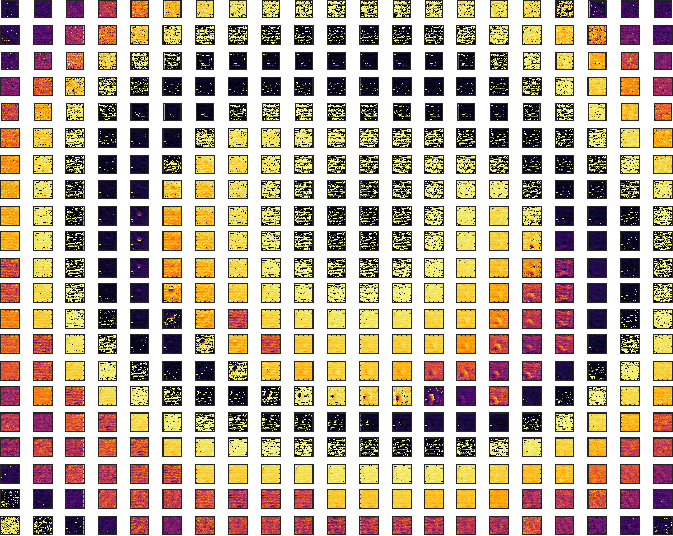}
          \caption{}
         \label{fig:intensity}
     \end{subfigure}
     \hfill
        \caption{The distribution of phase and intensity information across detector pixels. 300 nm polystyrene bead.}
        \label{fig:parallax_bead_comparision}
\end{figure}

The non-uniform capture of  information due to optical aberrations and diffraction effects is an idea that has been previously explored and exploited for increasing resolution and enhancing images \cite{rossman2021csparcom}. Specifically in \cite{rossman2021csparcom} they consider all aberrations within the mapping from object to image via an optimization algorithm, and do away with traditional pixel reassignment entirely. Characterizing the full system aberrations involves the direct measurements of the PSF by measuring a point-like object. \autoref{fig:parallax_bead_comparision} shows the intensity and phase response of the system to a 300 nm polymer bead. Specifically, the nonuniform, spherically abberated phase across the detector is emphasized. 
    \begin{figure}[h!]
     \centering
     \begin{subfigure}[b]{0.95\textwidth}
         \centering
         \includegraphics[width=\textwidth]{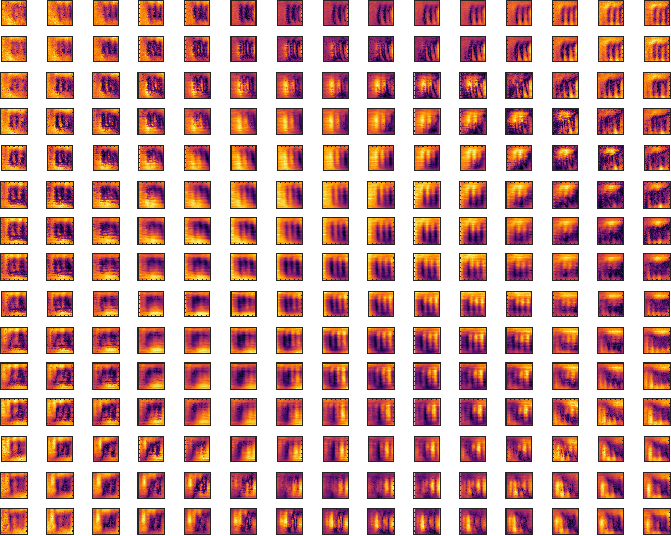}
                  \caption{}
         \label{fig:phase}
     \end{subfigure}
     \hfill
     \begin{subfigure}[b]{0.95\textwidth}
         \centering
         \includegraphics[width=\textwidth]{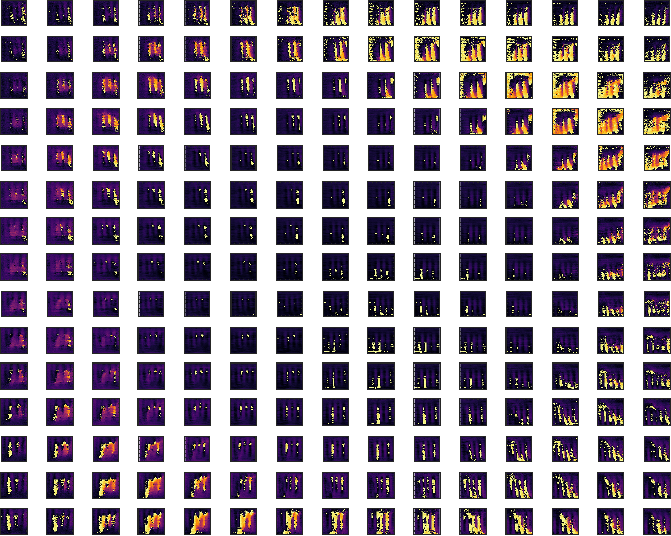}
                  \caption{}
         \label{fig:intensity}
     \end{subfigure}
     \hfill
     \end{figure}
\clearpage
         \begin{figure}
         \ContinuedFloat
          \centering
          \begin{subfigure}[b]{0.95\textwidth}
         \centering
         \includegraphics[width=1\textwidth]{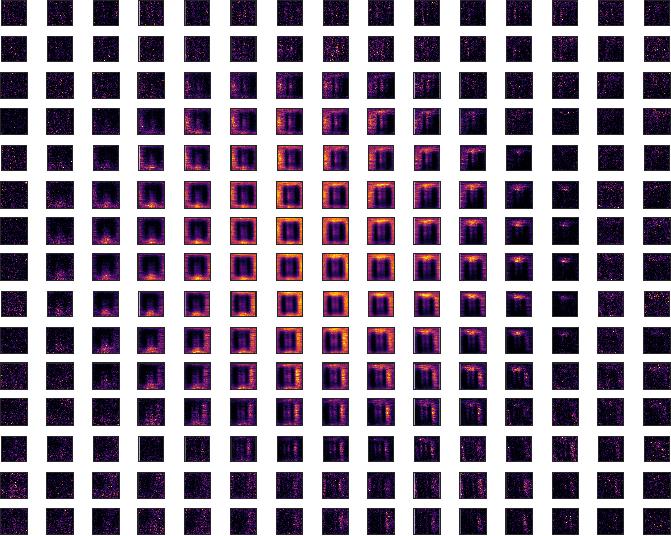}
                  \caption{}
         \label{fig:intensity}
     \end{subfigure}
     \hfill
        \caption{The distribution of a) intensity and b) phase information across detector pixels. c) shows basic CARS intensity images involving no field reconstruction (specifically no reference beam). 600 nm resolution target.}
        \label{fig:parallax_comparision}
\end{figure}

\clearpage

\section{Edge deconvolution}
\begin{figure}[h!]
    \centering
    \includegraphics[width=\textwidth]{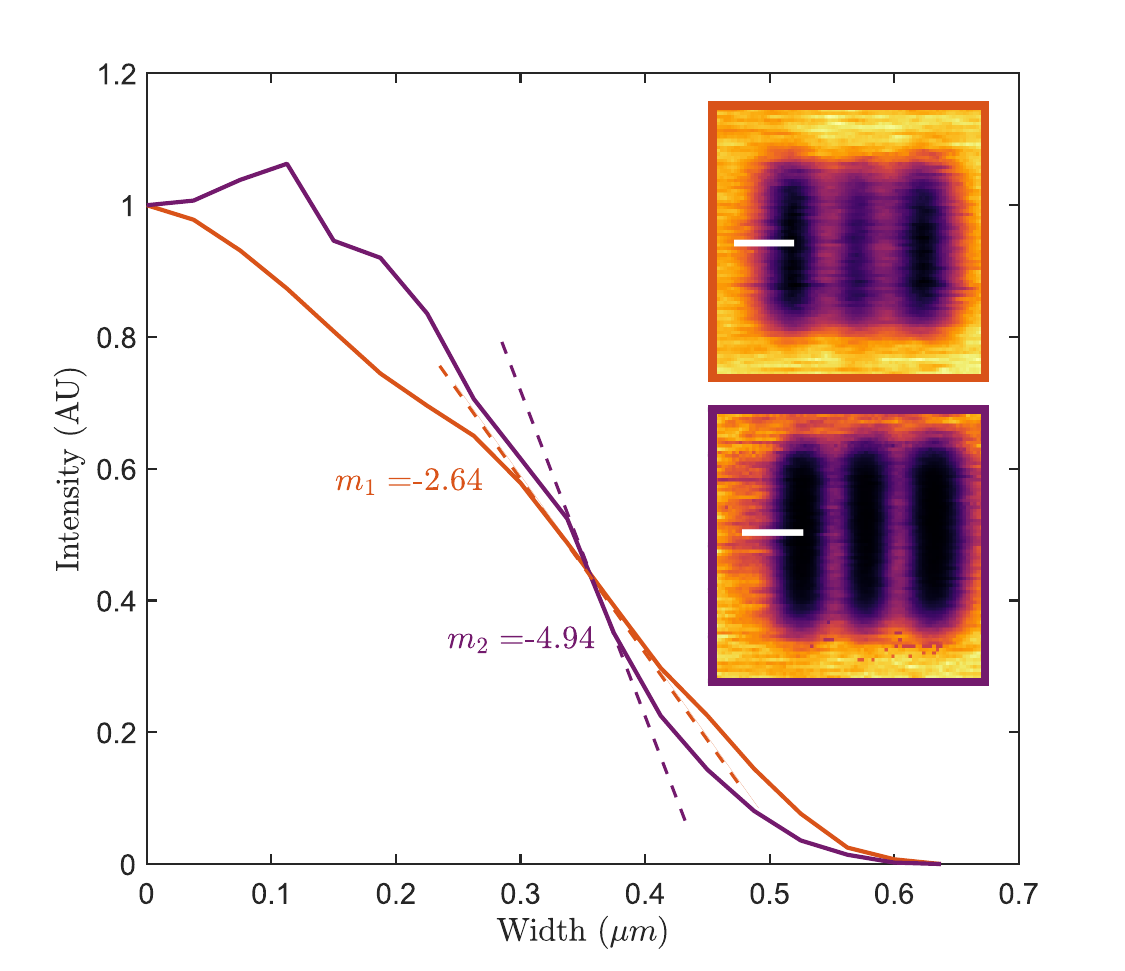}
    \caption{Edge deconvolution}
    \label{fig:edge_deconvolution}
\end{figure}

Quantification of the resolution enhancement is often performed through the width of the ASF by imaging a point like emitter. In the absence of a bright enough point like object for CARS, we turn to edge characterization. An analysis of line-cuts performed on the edges of the grooves in the 700 
 nmperiod target, is presented in \autoref{fig:edge_deconvolution}. The width of the edge profile - from the beginning of the inclination into the groove to the bottom, is less than a 100 nm which is much smaller than the expected width of the ASF, so that the edge profile can be approximated to a step function. Image of the edge is a convolution of the edge with the ASF, so we expect a line-cut to take the shape of a ``stretched out" step function. The degree of the stretching of the edge profile, indicative of the width of the ASF, can be characterized by the slope at the inflection point. Hence, the resolution enhancement factor can be estimated from the ratio of the slope of the line-cut of the cISM field amplitude image, and that of the open aperture intensity image, denoted $m_2$ and $m_1$ respectively. Our analysis yielded a resolution enhancement by a factor of $\Delta m \simeq 1.87$.

\clearpage

\section{Comparison with other super-resolution CARS techniques}

\begin{table}[h!]
\begin{tabular}{ |P{1.95cm}||P{1.2cm}|P{1.4cm}|P{1.5cm}|P{1.5cm}|P{1.4cm}| P{1.4cm}|P{1.4cm}|  }
 \hline
& Raman band & excitation power & basic resolution (nm) & achieved resolution (nm) & resolution gain & excitation NA & detection NA  \\
 \hline
 CARS-ISM& 2850 cm$^{-1}$ & 5 mW & 680 & 450  & 1.5-2 & 0.95    & 0.75 \\
  \hline
 CARS-RIM\cite{Fantuzzi2023}  & 2850 cm$^{-1}$ & 500 mW & 650 & 300  & 2.2 & 0.9    & 1.15 \\
  \hline
   FSRS-STED\cite{Silva2016}  & 1332 cm$^{-1}$ &  $10^{1}-10^{2}$ Wcm$^{-2}$ & 1350 & 820  & 1.65 & 0.4   & 0.55 \\
  \hline
SWM/EWM-CARS \cite{gong2020higher} & 2845 cm$^{-1}$ & $10^{12}$ Wcm$^{-2}$ & 328 & 230/196  & 1.5/1.7 & 1.1    & 1.4 \\
 \hline
 \end{tabular}
 \caption{Comparison of super-resolution CARS techniques.}
\label{table:comparison}
\end{table}

\end{document}